\documentclass[manuscript]{acmart}
\usepackage{enumitem}
\AtBeginDocument{%
  }

\copyrightyear{2026}
\acmYear{2026}
\setcopyright{cc}
\setcctype{by-nc-nd}
\acmConference[CHI '26]{Proceedings of the 2026 CHI Conference on Human Factors in Computing Systems}{April 13--17, 2026}{Barcelona, Spain}
\acmBooktitle{Proceedings of the 2026 CHI Conference on Human Factors in Computing Systems (CHI '26), April 13--17, 2026, Barcelona, Spain}
\acmPrice{}
\acmDOI{10.1145/3772318.3790515}
\acmISBN{979-8-4007-2278-3/2026/04}

\newcommand{\Modified}[1]{\textcolor{black}{#1}}

\begin{document}

\title{Understanding Nature Engagement Experiences of Blind People}

\author{Mengjie Tang}
\affiliation{%
  \institution{School of Computer Science and Engineering\\Southeast University}
  \city{Nanjing}
  \country{China}
}
\email{tangmengjie@seu.edu.cn}

\author{Xinman Li}
\affiliation{%
  \institution{School of Cyber Science and Engineering\\Southeast University}
  \city{Nanjing}
  \country{China}
}
\email{213222100@seu.edu.cn}

\author{Juxiao Zhang}
\affiliation{%
  \institution{Nanjing Normal University of Special Education}
  \city{Nanjing}
  \country{China}
}
\email{3301611@qq.com}

\author{Franklin Mingzhe Li}
\affiliation{%
  \institution{Carnegie Mellon University}
  \city{Pittsburgh}
  \state{Pennsylvania}
  \country{USA}
}
\email{mingzhe2@cs.cmu.edu}

\author{Zhuying Li}
\authornote{Corresponding author.}

\affiliation{%
  \institution{School of Computer Science and Engineering\\Southeast University}
  \city{Nanjing}
  \country{China}
}
\affiliation{%
  \institution{Key Laboratory of New Generation Artificial Intelligence Technology and Its Interdisciplinary Applications (Southeast University), Ministry of Education}
  \city{Nanjing}
  \country{China}
}
\email{zhuyingli@seu.edu.cn}
\renewcommand{\shortauthors}{Tang et al.}

\begin{abstract}
Nature plays a crucial role in human health and well-being, but little is known about how blind people experience and relate to it. We conducted a survey of nature relatedness with blind (N=20) and sighted (N=20) participants, along with in-depth interviews with 16 blind participants, to examine how blind people engage with nature and the factors shaping this engagement. Our survey results revealed lower levels of nature relatedness among blind participants compared to sighted peers. Our interview study further highlighted: 1) current practices and challenges of nature engagement, 2) attitudes and values that shape engagement, and 3) expectations for assistive technologies that support safe and meaningful engagement. We also provide design implications to guide future technologies that support nature engagement for blind people. Overall, our findings illustrate how blind people experience nature beyond vision and lay a foundation for technologies that support inclusive nature engagement.
\end{abstract}

\ccsdesc[500]{Human-centered computing~Empirical studies in accessibility}

\keywords{Nature engagement, Blind people, Accessibility, Assistive technology}


\maketitle

\section{Introduction}
Nature is a vital resource for human wellbeing, offering benefits such as stress reduction, emotional regulation, and psychological restoration \cite{howard2017nature, ROMAGOSA201570, PROBSTLHAIDER20151, kondo2020nature, shanahan2019nature}. This recognition has led to increasing interest in Human–Nature Interaction (HNI), which examines how people perceive, access, and engage with natural environments \cite{webber2023engaging, spors2023longing, wu2025greencompass}. However, existing HNI research has predominantly focused on sighted people. For blind people, nature engagement is equally important, contributing to their sense of belonging, identity, and psychological recovery \cite{shaw2015exploring, bandukda2019understanding}. Unfortunately, most nature-based experiences are designed with visual perception in mind \cite{packer2008tourist, chikuta2019accessibility, wall2023access}. As a result, blind people often rely on sighted companions to interpret their surroundings, which limits their autonomy, independence, and the depth of their engagement with nature \cite{bandukda2019understanding, groulx2022accessible, bandukda2020places,li2023understanding}.

Efforts to improve accessibility in parks and nature reserves have introduced physical infrastructure such as tactile paving, tactile maps, and Braille signage, enhancing mobility for blind visitors \cite{npsBlindAccess2024, kuriakose2022tools,li2021choose, siu_accessible_2013}. Additionally, prior research has developed digital solutions like voice-based navigation systems \cite{harsur2017voice, cha2013design}, ambient sound feedback \cite{bandukda2020audio}, and multisensory augmented reality interfaces that combine tactile and audio feedback to enhance spatial information access \cite{albouys2018towards}. However, these solutions remain predominantly function-oriented, focusing on wayfinding and spatial orientation, while offering limited support for the experiential dimensions of nature engagement, such as sensory discovery, emotional resonance, and social interaction \cite{helen1997british, hoogsteen2022beyond, india2021vstroll}. To better support blind people's full engagement with nature, a detailed understanding of their sensory strategies, barriers, and the type of experiential support needed from assistive technologies is essential.

Our study addresses the following research questions:
\begin{itemize}[topsep=0pt, partopsep=0pt]
  \item RQ1: How do blind people engage with nature and what challenges do they face? 
  \item RQ2: What expectations do blind people express for assistive technologies that could support nature engagement?
  \item RQ3: How can technologies be designed to support inclusive, autonomous, and emotionally rich nature engagement for blind people?
\end{itemize}

To answer these questions, we conducted a comparative survey of the Nature Relatedness Scale (NRS) \cite{nisbet2009nature} with blind (N=20) and sighted (N=20) participants, followed by semi-structured interviews with 16 blind participants. The survey revealed lower NRS scores among blind participants, suggesting constraints in their affective and cognitive connectedness to nature. The interview findings provided a detailed account of how blind people engage with nature through multisensory strategies, typically within safe, familiar, and socially supported environments (Section~\ref{experiences and challenges}). We also uncovered how participants attributed emotional value to nature, expressed varied attitudes toward nature engagement, and articulated degrees of environmental awareness that shape their engagement (Section~\ref{attitudes}). Building on these insights, our participants described their expectations for assistive technologies in nature engagement, emphasizing systems that improve access to environmental information, create immersive and emotionally resonant experiences, enable accessible ways of recording and recalling encounters, and provide support that is portable and sensitive to different nature contexts (Section~\ref{expectation}). Finally, we discuss these results in relation to existing work, and outline design implications for technologies that move beyond mobility support toward richer, multisensory, and socially connected engagement with nature (Section~\ref{discussion}). Overall, we believe our findings and discussion contribute opportunities to support blind people in leveraging nature-related technologies for safer and richer engagement with the natural world.

\section{Related Work}
\label{related work}
In this section, we first highlight the striking lack of research on how blind people can engage with nature, identifying this as the core gap our work addresses (Section \ref{The importance of nature engagement for blind people}). We then examine potential assistive technologies that could support such engagement (Section \ref{Potential Assistive Technologies}). Finally, we situate these discussions within the broader landscape of technology-mediated nature engagement research (Section \ref{sec:theoretical_foundations}).

\subsection{Nature Engagement for Blind People}

\label{The importance of nature engagement for blind people}
A substantial body of research demonstrates that natural environments play an important role in promoting human health and well-being \cite{howard2017nature, ROMAGOSA201570, PROBSTLHAIDER20151, jimenez2021associations, kondo2020nature}. Viewing natural landscapes reduces stress and aids recovery \cite{ulrich1984view}, activities in green spaces restore attention and relieve fatigue \cite{kaplan1995restorative}, and contact with nature improves mood and lowers depression risk \cite{bratman2019nature, shanahan2019nature}. For blind people, nature engagement is equally vital, supporting psychological restoration, emotional well-being, and identity formation \cite{shaw2015exploring, bandukda2019understanding}.

To address accessibility, parks and reserves have introduced tactile paving, tactile maps, and Braille signage to enhance spatial access \cite{npsBlindAccess2024, kuriakose2022tools, li2021choose, siu_accessible_2013}. Research has also proposed outdoor navigation and path-planning technologies \cite{golledge1996cognitive, schinazi2016spatial, loomis2005personal}. While these initiatives improve mobility, they give limited support to information access, emotional regulation, and immersive experience, leaving a gap between functional accessibility and experiential inclusion.

Nature engagement extends beyond physical presence to multisensory and affective experiences \cite{shaw2015exploring, bandukda2019understanding}. Blind people actively participate through compensatory mechanisms that amplify non-visual modalities \cite{bell2019cross, franco2017benefits,li2022feels}. Cues such as wind, birdsong, flowing water, and tactile exploration of plants foster spatial understanding and attachment, with nature often described as a “restorative refuge” \cite{shaw2015exploring, bandukda2019understanding}. Natural soundscapes further reduce stress and aid recovery even without vision \cite{alvarsson2010stress, ratcliffe2021sound, li2024influence}, while cross-modal plasticity enhances auditory and tactile immersion \cite{bell2019cross, franco2017benefits}. Multisensory cues such as sound, smell, and touch also support participation and safety in public spaces \cite{jenkins2015experience}. These findings underline the experiential dimensions essential to inclusion.

Yet built environments remain predominantly vision-centric. Heavy reliance on visual signage and cues undermines autonomy and immersion for blind people \cite{packer2008tourist, chikuta2019accessibility, wall2023access, jenkins2015experience}. Such limitations constrain everyday experiences and reflect a lack of systematic understanding of blind people’s engagement with nature. By examining their sensory practices and challenges, our study sheds light on how blind people construct meaning through non-visual interactions with natural environments, offering empirical insights into non-visual experiences of nature engagement.

\subsection{Potential Assistive Technologies for Blind People’ Nature Engagement}
\label{Potential Assistive Technologies}
Existing HCI and accessibility research shows that assistive technologies have the potential to support blind people’s engagement with nature \cite{bigham2010vizwiz, wang2021toward}. These efforts can broadly be divided into two categories: functionality-oriented designs, which emphasize mobility and information access \cite{ahmetovic2016navcog, golledge1996cognitive, schinazi2016spatial}, and experience-oriented designs, which highlight the role of multisensory and affective approaches in creating immersion and resonance \cite{li2023understanding, chang2022omniscribe, bruns2023touch}.

In functionality-oriented research, system design has mainly focused on information access and navigation, a direction that has long dominated accessibility and spatial cognition studies \cite{golledge1996cognitive, schinazi2016spatial,li2025oscar,li2025exploring,ning2025aroma, loomis2005personal,li2024contextual,kianpisheh2019face}. Information access tools such as VizWiz \cite{bigham2010vizwiz} and Be My Eyes \cite{bemyeyes} enable users to obtain real-time scene descriptions via crowdsourcing or remote assistance. Navigation and path-planning systems, including the NavCog series \cite{ahmetovic2016navcog, sato2017navcog3} and haptic feedback devices \cite{velazquez2018outdoor, kammoun2012guiding}, provide spatial guidance and terrain awareness. More recently, AI systems such as WorldScribe and VocalEyes \cite{chang2024worldscribe, chavan2024vocaleyes} have demonstrated potential in contextual environmental narration, while VIPTour and ChatMap \cite{lin2025ai, hao2024chatmap} explore personalized navigation and cognitive mapping. Collectively, these technologies significantly enhance mobility, safety, and task efficiency, yet their design logic remains primarily task-oriented, with limited attention to immersive and emotional dimensions \cite{loomis2005personal}.

In contrast, experience-oriented research emphasizes the role of multisensory design in fostering richer engagement \cite{welewatta2024sema,li2023understanding, cho2021study, li2022feels, li2025more, li2021non}. Audio and narrative work has shown that environmental sound, when combined with narration or enhanced description, can create stronger presence and atmosphere \cite{wang2021toward, walczak2018audio, rector2017eyes}. Spatialized and co-created narratives in 360° video and VR performance further illustrate how sound can be designed for immersion rather than just information delivery \cite{chang2022omniscribe, natalie2024audio, jiang2023beyond, dang2024towards}. Beyond audio, cross-modal approaches demonstrate how pairing touch with hearing can evoke emotional resonance in artistic settings \cite{li2023understanding}, while tactile artifacts such as 3D-printed models foster cultural connection and identity \cite{bruns2023touch}. Extending further, multisensory explorations that combine modalities like sound, temperature, and tactile augmentation show how aesthetic and affective expression can be broadened \cite{cho2021study, bartolome2021multi, welewatta2024sema}. However, most of these explorations remain confined to artistic and cultural domains, with systematic applications in nature engagement still largely absent.

Overall, while existing technologies have achieved progress in both functionality and experience, functionality-focused solutions often remain overly task-driven, and experiential approaches are largely limited to art and virtual contexts. Systematic integration of the two in natural settings is still lacking. Therefore, further research is needed to bridge this gap and shift technological design from merely ``ensuring mobility'' toward ``enriching experience,'' ultimately creating more comprehensive support for blind people’s engagement with nature. Rather than focusing solely on functional access, our study highlights the value of designing technologies that support rich, situated experiences grounded in blind people’ sensory and emotional relationships with nature, and offers a foundation for future inclusive, experience-centered design.

\subsection{Technology-Mediated Nature Engagement}
\label{sec:theoretical_foundations}
Human–Nature Interaction (HNI) has gradually evolved into a design-oriented research domain within HCI, grounded in diverse theoretical traditions. Early perspectives from environmental psychology and biophilia highlighted the restorative and affiliative value of nature \cite{kaplan1995restorative, kellert1993biophilia}. More recent HNI research, however, is informed by critical and design-oriented theories that reframe human–nature relations. Posthumanism challenges anthropocentrism by emphasizing the agency of nonhuman life \cite{braidotti2013posthuman, forlano2017posthumanism}; Wakkary’s notion of “more-than-human design” positions artifacts and technologies as active participants in ecosystems \cite{wakkary2021things}; Liu et al.’s concept of “co-habiting survival” calls for design that fosters coexistence and shared flourishing across species \cite{liu2018design}; and Altarriba Bertran et al. draw on the idea of the “playfulness of forests” to argue for embedding joy, embodiment, and symbolic meaning in nature–technology interactions \cite{altarriba2023playful}. Together, these perspectives reposition HNI from a human-centered focus toward multispecies interdependence \cite{chen2019ipanda, li2022re}, ecological care \cite{maher2014naturenet, kawas2020otter}, and new possibilities for technological mediation of nature engagement \cite{stepanova2020jel, mostajeran2023effects}.  

In practice, some research emphasizes how technology can support knowledge acquisition \cite{ibird, maher2014naturenet}, task guidance \cite{wikiloc2022, pielot2012tacticycle, fedosov2016skiar}, and active participation \cite{rogers2004ambient, kawas2020otter, gaver2019my} in ecosystems. Navigation systems embed environmental awareness into outdoor practices, such as hiking or cycling, where Wikiloc enables users to record routes and develop spatial understanding \cite{wikiloc2022}, while Tacticycle and SkiAR employ haptic feedback and augmented reality to enhance perception of environmental cues \cite{pielot2012tacticycle, fedosov2016skiar}. Educational systems lower barriers to learning through interactive or gamified experiences: Ambient Wood engaged children in woodland inquiry using sensor-based triggers \cite{rogers2004ambient}; iBird provided interactive bird identification to enhance everyday ecological literacy \cite{ibird}; and IPANDA combined physical interaction with digital play to raise awareness of wildlife protection \cite{chen2019ipanda}. Sustained ecological action has also been supported through platforms and community-based tools. For example, NatureNet encouraged users to share ecological data and participate in citizen science \cite{maher2014naturenet}, NatureCollections fostered children’s everyday engagement by gamifying photo collection of natural elements \cite{kawas2020otter}, and MyNatureWatch enabled urban residents to document wildlife with DIY cameras, sparking long-term ecological awareness \cite{gaver2019my}.  

Other work focuses on multisensory and embodied design to evoke affective resonance and symbolic connection with nature \cite{stepanova2020jel, li2022re, mostajeran2023effects, li2025tuning, xu2024istraypaws}. The JeL project, for instance, engages users in synchronized breathing and movement to cultivate a felt sense of connection with natural rhythms \cite{stepanova2020jel}. GoChirp detects birdsong and translates it into haptic feedback, creating everyday encounters with wildlife and evoking surprise and intimacy \cite{li2022re}. Virtual nature systems have also demonstrated the potential of immersive environments for psychological restoration, where virtual forests, waterscapes, and multisensory atmospheres can enhance positive emotions and foster recovery \cite{mostajeran2023effects, kalantari2022using, yu2020restorative, feng2018closer, van2021virtual}. Collectively, these studies highlight the role of the body, the senses, and emotions in shaping nature experiences, broadening the scope of technological mediation in human–nature connectedness.  

Despite these advances, most existing work remains centered on sighted populations. Designs often rely on visual presentation, with limited attention to the potential of non-visual modalities such as touch and hearing. As a result, blind people continue to face barriers to equitable and rich engagement with nature. Our study addresses this gap by drawing on empirical interviews to surface the sensory practices and challenges of blind people, and by offering design implications that inform more inclusive, embodied, and emotionally resonant forms of human–nature interaction.

\section{Survey on Nature Relatedness in Blind People}
\label{NRS}
To provide a comparative baseline for our qualitative study, we conducted a pre-registered survey to examine differences in nature relatedness between blind and sighted participants. 

\begin{table*}[h]
\caption{Descriptions and examples of the three factors derived from the NRS.}
\Description{This table summarizes the three subscales of the Nature Relatedness Scale. NR-Self refers to internal identification with nature, NR-Perspective to a worldview about humans' impact on nature, and NR-Experience to physical comfort and desire for being outdoors.}
\label{nrs_three}
\centering
\begin{tabular}{p{2cm}p{7cm}p{5cm}}
\toprule
\textbf{Factor} & \textbf{Description} & \textbf{Example Item} \\
\midrule
NR-Self & An internalized identification with nature, reflecting feelings and thoughts about one’s personal connection to nature. & ``My connection to nature and the environment is a part of my spirituality.'' \\
NR-Perspective & An external, nature-related worldview, a sense of agency concerning individual human actions and their impact on all living things. & ``The state of nonhuman species is an indicator of the future for humans.'' \\
NR-Experience & A physical familiarity with the natural world, the level of comfort with and desire to be out in nature. & ``I enjoy being outdoors, even in unpleasant weather.'' \\
\bottomrule
\end{tabular}
\end{table*}

\subsection{Method}

\subsubsection{Survey Recruitment}
The blind group (N=20) was recruited through the local blind association, and the sighted group (N=20) was recruited online via social media groups (e.g., QQ and WeChat\footnote{Widely used mobile messaging and social media applications in China \cite{cnnic55,tencent_q4_2024}.}). 
To participate in our survey study, all participants were required to be between 18 and 60 years old, with blind participants further required to be either legally blind or totally blind. 
Each session was administered via Tencent Questionnaire\footnote{An online survey platform in China, similar to Google Forms \cite{tencent_wj_about}.} and lasted approximately 5 minutes. 
Participants were compensated with 5 CNY (0.7 USD). 
All recruitment and study procedures were reviewed and approved by the local ethics committee.

\subsubsection{Survey Questions}
In addition to the main survey scale, we first collected basic demographic information, including age, gender, vision level, and onset age of blindness. 
We then employed the 21-item Nature Relatedness Scale (NRS) \cite{nisbet2009nature}, a validated psychometric tool developed to measure affective, cognitive, and experiential aspects of individuals’ connection to nature. The scale consists of 21 items rated on a 5-point Likert scale (1 = strongly disagree, 5 = strongly agree), with eight items reverse-scored. These items are grouped into three factors: NR-Self, NR-Perspective, and NR-Experience (Table~\ref{nrs_three}).

\subsubsection{Data and Analysis}
In total, we initially received 24 responses from blind participants and 52 from sighted participants. 
After removing duplicates, incomplete, or ineligible entries, 4 blind and 7 sighted responses were excluded. 
To ensure demographic comparability, we applied frequency matching on age range and gender distribution, resulting in a balanced dataset of 20 blind and 20 sighted participants for analysis. 
The final blind sample included 20 participants: 10 self-identified as male and 10 as female, with a mean age of 37.2 years ($SD = 13.5$). Sixteen participants reported congenital blindness and four reported acquired blindness. The sighted sample also included 20 participants: 10 self-identified as male and 10 as female, with a mean age of 37.4 years ($SD = 14.4$). All participants were given the option to identify as male, female, nonbinary, self-described, or prefer not to say; none selected the latter three categories.

Given the limited sample size and the ordinal nature of Likert-scale data, group differences were assessed using Mann--Whitney $U$ tests as an exploratory nonparametric approach.

\subsection{Results}
\label{results:nrs}
\begin{figure}[t!]
    \centering
    \includegraphics[width=\linewidth, trim=50 480 50 50, clip]{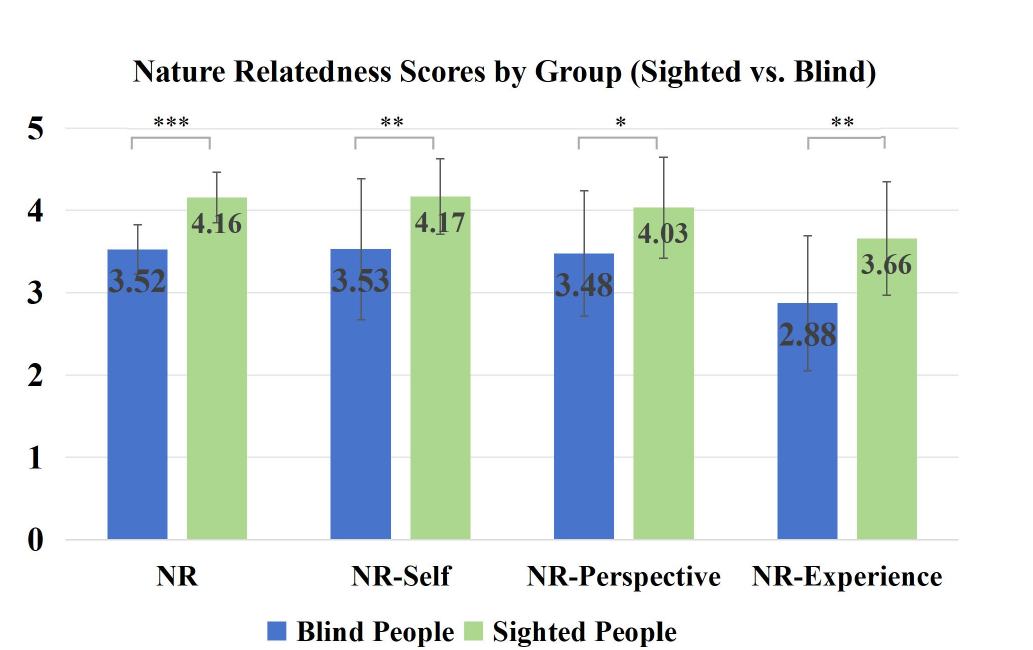}
    \caption{Group differences in NRS subscales and the overall NR score. Error bars show standard deviations. Asterisks indicate significant group differences (Mann--Whitney U tests; * $p < .05$, ** $p < .01$, *** $p < .001$).}
    \label{fig:nrs_summary}
    \Description{Bar chart comparing blind and sighted participants on the Nature Relatedness Scale. Sighted participants scored higher on the overall NRS and on each subscale, including NR-Self, NR-Perspective, and NR-Experience. Blind participants consistently scored lower. Error bars indicate variability (mean $\pm$ SD). Statistically significant group differences are marked by asterisks: one asterisk for $p < .05$, two for $p < .01$, and three for $p < .001$.}
\end{figure}

Group comparisons of Nature Relatedness Scale (NRS) scores between sighted and blind participants are shown in Figure~\ref{fig:nrs_summary} and detailed in Table~\ref{tab:nrs_comparison}. Using two-tailed Mann--Whitney $U$ tests, we found the following. For the overall NR score, blind participants scored lower than sighted participants ($U=34.5$, $p<.001$, $r=.71$), indicating a large effect. For NR-Self, scores were lower among blind participants ($U=93.0$, $p=.003$, $r=.46$), a medium--large effect suggesting weaker incorporation of nature into self-concept and feelings. For NR-Perspective, blind participants also scored lower ($U=109.5$, $p=.014$, $r=.39$), a medium effect reflecting differences in environmental worldview and perceived agency. For NR-Experience, a significant difference again favored sighted participants ($U=91.0$, $p=.003$, $r=.47$), a medium--large effect consistent with reduced comfort or ease of direct engagement in natural settings. Effect magnitudes followed the pattern: overall NR (large) $>$ NR-Experience $\approx$ NR-Self (both medium--large) $>$ NR-Perspective (medium).

These results show consistently higher scores among sighted participants across the overall NR score and all three subscales, with the largest disparity observed in the overall NR score, followed by NR-Self and NR-Experience, and a smaller but still significant difference in NR-Perspective. These differences suggest that visual experience plays a key role in shaping not only sensory familiarity but also environmental attitudes and identity integration. 

\begin{table}[h]
\caption{Comparison of NRS subscales and the overall NR score between blind and sighted participants ($r$ as effect size).}
\Description{This table compares blind and sighted participants on the Nature Relatedness Scale. Across all subscales and the overall score, blind participants scored significantly lower. Effect sizes were large for the overall NR score, medium to large for NR-Self and NR-Experience, and medium for NR-Perspective.}
\label{tab:nrs_comparison}
\centering
\begin{tabular}{lcccc}
\toprule
\textbf{Dimension} & \textbf{$U$} & \textbf{$p$-value} & \textbf{$r$} & \textbf{Interpretation} \\
\midrule
NR & 34.5  & $<.001$ & 0.71 & Large \\
NR-Self          & 93.0 & 0.003  & 0.46 &  Medium--Large \\
NR-Perspective   & 109.5  & 0.014  & 0.39 & Medium \\
NR-Experience    & 91.0  & 0.003  & 0.47 & Medium--Large \\
\bottomrule
\end{tabular}
\end{table}

At the item level, group differences emerged across the full set of 21 NRS items after reverse scoring where appropriate (Appendix A, Figure~\ref{fig:nrs_appendix}). Seven of the 21 items reached nominal significance ($p < .05$). Among these, six showed at least a moderate effect ($r \ge .30$) and are highlighted below; the remaining significant item exhibited a small effect ($r < .30$) and is reported in Appendix A, Table~\ref{tab:nrs_items}. The six items were: \textit{``I enjoy being outdoors, even in unpleasant weather''} (Item~1); \textit{``I enjoy digging in the earth and getting dirt on my hands''} (Item~6); \textit{``I take notice of wildlife wherever I am''} (Item~9); \textit{``Nothing I do will change problems in other places on the planet''} (Item~11); \textit{``Conservation is unnecessary because nature is strong enough to recover from any human impact''} (Item~18); and \textit{``I think a lot about the suffering of animals''} (Item~20).

Viewed through the three NRS dimensions, the six significant items form a coherent pattern. In NR-Experience, Items~1 ($U=118$, $p=.025$, $r=.35$), 6 ($U=115$, $p=.018$, $r=.36$), and 9 ($U=105$, $p=.008$, $r=.41$) indicate lower outdoor enjoyment, reduced tactile engagement, and fewer incidental notices of wildlife among blind participants, pointing to constrained opportunities for spontaneous multisensory contact with nature. In NR-Perspective, Items~11 ($U=107$, $p=.009$, $r=.40$) and 18 ($U=75$, $p<.001$, $r=.54$) reflect reduced perceived agency and greater belief that nature can recover without conservation. In NR-Self, Item~20 ($U=114$, $p=.015$, $r=.37$) shows lower empathic concern for nonhuman life, suggesting a weaker incorporation of concern for the natural world and nonhuman life into self-relevant feelings and values. Taken together, the three dimensions trace a linked pathway from constrained experience to attenuated self-integration and diminished environmental agency.

Overall, the results reveal a consistent disadvantage for blind participants across both affective and cognitive dimensions of nature relatedness. The absence of visual perception appears to limit spontaneous engagement with natural environments, thereby reducing experiential comfort, ecological worldview formation, and the integration of nature into self-identity. While non-visual sensory pathways provide partial access, they may not fully substitute for the integrative perceptual and symbolic roles of vision \cite{loomis2018sensory}.

\section{Semi-structured Interview}
\label{Semi-structured Interview}
We conducted semi-structured interviews to understand blind people’ experiences with nature, the challenges they face, and their expectations for assistive technologies in supporting nature engagement. 

\subsection{Participants}
We recruited 16 blind participants (P1–P16) for interviews from the same group who had participated in the earlier survey (Table \ref{table:participants}). These individuals were selected based on their availability and willingness to participate in a follow-up session. Among the 16 participants, 10 self-identified as male, 6 as female, and none selected nonbinary, self-described, or preferred not to report gender (Table~\ref{table:participants}). The average age was 41.3 years ($SD = 12.0$). Eight participants were legally blind, while eight were completely blind. Eleven participants had congenital blindness, while the remainder had lost vision at an average age of 13. Each interview lasted approximately 45 to 60 minutes. Participants received a compensation of 50 CNY (7 USD). The recruitment and research procedures were approved by the local ethics committee.

\def\arraystretch{1.15}
\begin{table*}[t]
\caption{Participants' demographic information \Modified{and nature engagement context (All residences are cities in China)}.}
\Description{This table presents demographic and contextual information for 16 participants. It contains eight columns: ID, Gender, Age, Blindness, Occupation, Living Situation, Residence, and Nature Access.}
\label{table:participants}
\centering
\resizebox{1\textwidth}{!}{%
\begin{tabular}{
p{0.4cm}|
p{0.9cm}
p{0.4cm}
p{2.8cm}
p{2.3cm}
p{2.8cm}
p{1.5cm}
p{3.0cm}
}
\toprule
\textbf{ID} & 
\textbf{Gender} &
\textbf{Age} & 
\textbf{Blindness} & 
\textbf{Occupation} &
\Modified{\textbf{Living Situation}} & 
\Modified{\textbf{Residence}} &
\Modified{\textbf{Nature Access}} \\
\midrule

P1  & Male & 22 & Legally blind (congenital) & Student &
\Modified{With family or friends} &
\Modified{Nanjing (Suburban)} &
\Modified{Nearby parks and other small greenspaces.} \\

P2  & Male & 25 & Totally blind (congenital) & Massage therapist &
\Modified{With family or friends} &
\Modified{Shaoyang (Suburban)} &
\Modified{Nearby parks and accessible hilly natural areas.} \\

P3  & Female & 22 & Totally blind (congenital) & Student &
\Modified{With family or friends} &
\Modified{Nanjing (Urban)} &
\Modified{Campus gardens and small greenspaces directly accessible.} \\

P4  & Female & 36 & Totally blind (acquired at age 14) & Unemployed &
\Modified{With family or friends} &
\Modified{Shijiazhuang (Urban)} &
\Modified{Nearby parks and other small greenspaces.} \\

P5  & Male & 40 & Legally blind (acquired in 2011) & Massage therapist &
\Modified{Independent} &
\Modified{Shijiazhuang (Urban)} &
\Modified{Parks near metro-accessible areas.} \\

P6  & Male & 47 & Totally blind (congenital; became totally blind at age 30) & Massage therapist &
\Modified{With family or friends} &
\Modified{Shijiazhuang (Urban)} &
\Modified{Nearby parks and other small greenspaces.} \\

P7  & Male & 63 & Totally blind (acquired early; fully blind mid-life) & Massage therapist &
\Modified{With family or friends} &
\Modified{Shijiazhuang (Suburban)} &
\Modified{Nearby parks and other small greenspaces.} \\

P8  & Female & 37 & Legally blind (congenital; light perception only) & Massage therapist &
\Modified{With family or friends} &
\Modified{Shijiazhuang (Urban)} &
\Modified{Nearby parks and other small greenspaces.} \\

P9  & Female & 37 & Legally blind (congenital) & Massage therapist &
\Modified{With family or friends} &
\Modified{Shijiazhuang (Urban)} &
\Modified{Nearby parks and other small greenspaces.} \\

P10 & Female & 42 & Legally blind (acquired in 2006) & Unemployed &
\Modified{With family or friends} &
\Modified{Shijiazhuang (Urban)} &
\Modified{Nearby parks and other small greenspaces.} \\

P11 & Female & 43 & Legally blind (congenital; light perception only) & Massage therapist &
\Modified{With family or friends} &
\Modified{Shijiazhuang (Urban)} &
\Modified{Nearby parks and other small greenspaces.} \\

P12 & Male & 40 & Totally blind (congenital) & Massage therapist &
\Modified{With family or friends} &
\Modified{Cangzhou (Rural)} &
\Modified{Nearby rural nature areas; parks and mountains also accessible.} \\

P13 & Male & 50 & Legally blind (congenital) & Massage therapist &
\Modified{Independent} &
\Modified{Shijiazhuang (Urban)} &
\Modified{Nearby parks and other small greenspaces.} \\

P14 & Male & 42 & Totally blind (congenital) & Self-employed &
\Modified{With family or friends} &
\Modified{Shijiazhuang (Urban)} &
\Modified{Nearby parks and other small greenspaces.} \\

P15 & Male & 57 & Totally blind (acquired at age 1) & Massage therapist &
\Modified{With family or friends} &
\Modified{Dingzhou (Suburban)} &
\Modified{Nearby parks and other small greenspaces.} \\

P16 & Male & 57 & Totally blind (congenital; became totally blind two years ago) & Officer &
\Modified{With family or friends} &
\Modified{Chengde (Urban)} &
\Modified{Nearby parks and other small greenspaces, with the Mountain Resort accessible on foot.} \\

\bottomrule
\end{tabular}%
}
\end{table*}

\subsection{Interview Procedure}
Each semi-structured interview began with a brief introduction to the study’s purpose, followed by a short warm-up segment, and then the collection of participants’ self-reported demographic information (e.g., age, gender, and level of visual impairment) to contextualize participant responses.

\subsubsection{Nature Engagement Experiences and Challenges:}
We then asked about participants' experiences with nature, such as the amount of time they typically spend in natural settings each week, the forms of engagement, companions during these activities, memorable experiences, and any challenges encountered. Beyond overall experiences, we also inquired about preferred natural elements and the degree of sensory reliance during engagement.

\subsubsection{Preferences and Expectations for Assistive Technologies in Nature Engagement:}
Participants were asked about assistive technologies they had used while engaging with nature and their experiences with them. We also introduced additional existing technologies not mentioned by participants and discussed their preferences and expectations for future tools.

\subsubsection{Attitudes Toward Nature Engagement:}
Lastly, we explored participants' attitudes toward nature engagement. For those with acquired blindness, we asked them to reflect on any shifts in their engagement with nature before and after vision loss to understand the influence of visual deprivation on their motivation.

All semi-structured interviews were conducted via Tencent Meeting\footnote{An online video conferencing platform in China, similar to Zoom.}, audio-recorded, and transcribed verbatim. \Modified{The interview protocol is provided in Appendix B}.

\subsection{Data Analysis}
\Modified{We conducted an inductive thematic analysis of the interview data \cite{braun2006using}.}
One researcher first transcribed the interviews verbatim into text. Two researchers then independently reviewed the full interview transcripts and performed open coding \cite{charmaz2006constructing}. \Modified{After initial coding, the two researchers compared and discussed their codes to clarify overlapping concepts and develop a shared interpretive understanding. They then collaboratively clustered related codes and developed higher-level themes.} A total of 64 initial codes were generated during open coding. Through discussion and consolidation, \Modified{we developed these into 10 focused codes and interpretively organized them into three overarching themes.} Each theme was iteratively refined to ensure interpretative clarity and thematic representativeness.

\section{\Modified{Findings: Nature Engagement Is Structured by Safety, Limited Environmental Cues, and Multisensory Strategies}}
\label{experiences and challenges}
\Modified{This section shows how blind people’s nature engagement forms through a layered process. Safety functions as the foundational condition, shaping when and where engagement is possible. On this basis, limited situational cues constrain active exploration, reducing opportunities for spontaneous discovery. Despite these constraints, participants build connection through multisensory strategies, using hearing, smell, and touch to interpret and experience natural environments. Together, these dimensions outline how access, exploration, and sensory engagement jointly shape nature encounters.}

\subsection{\Modified{Safety as the Precondition for Nature Engagement}}
\label{safe}
For blind people, engaging with nature often began with a fundamental consideration: whether it felt safe to do so. Participants’ choices of where to go, how to move through space, and whether to be accompanied were all deeply shaped by concerns around risk and security.

A strong preference for \textbf{familiar and low-risk environments} shaped ten participants’ outdoor routines. Ten participants often engaged with nature in nearby green spaces or community parks, where environments were more predictable and manageable. P5 explained, \textit{``I avoid unfamiliar places entirely... The routes I take are basically fixed, and since I’ve walked them many times, there are generally no problems.''} Similarly, P4 noted, \textit{``In familiar environments, I can sense the surroundings, the nearby buildings... but in unfamiliar places, that ability drops.''} Such accounts underscore how prior exposure and repetition reduced risks and supported navigation. For participants with congenital blindness, the absence of early-life reference points could make unfamiliar environments feel even more inaccessible. As P13 put it, \textit{``Some people just don't know what the outside world is like... so they don't go out.''} Nature engagement, in these cases, was tightly coupled with spatial familiarity as a condition for safe movement.

\textbf{Companionship} was regarded as another crucial form of safety support. Ten participants described relying on companions as a condition for going outdoors. Rather than convenience, accompaniment served a functional role in reducing risk and enabling engagement. P4 shared, \textit{``I almost never go out on my own... I usually stay at home unless my husband or parents take me out.''} Likewise, P7 emphasized, \textit{``No matter what plans I have, I need to go with friends or family. Even a group of blind people wouldn’t go without someone sighted.''} These accounts illustrate how accompaniment often played an indispensable role in safe access to nature.

\textbf{Family restrictions} also operated as a form of safety enforcement. Families, while motivated by care, often discouraged independent outings in ways that reduced autonomy and limited opportunities for independent practice. P4 explained, \textit{``I really want the ability to go out on my own, but my parents always say there are too many cars, and they worry too much.''} P11 added, \textit{``If your family thinks you should just stay at home to avoid accidents, you lose the courage to go out.''} These protective attitudes, though well-intentioned, could restrict engagement by reinforcing dependency.

In sum, blind people’s access to nature is structured not only by physical accessibility, but by a network of safety-related factors: the reliance on familiar environments, the importance of accompaniment, and the constraints of family protection. These findings suggest that while independent mobility technologies are vital, safety also depends on social and environmental dynamics, which form the foundation for rich and secure engagement with natural spaces.

\subsection{\Modified{Limited Situational Cues Constrain Active Exploration}}
\label{Active Exploration}

Our participants’ engagement with nature was shaped not only by safety needs but also by how exploration unfolded in practice. Despite strong sensory capacities and curiosity, they often lacked \Modified{\textbf{real-time situational cues that support spontaneous exploration}}. As P7 noted, \textit{``Sighted people may suddenly turn right because something catches their eye, but blind people just keep walking straight.''} Without incidental visual stimuli, their movements became linear and predetermined, limiting the possibility of unexpected encounters.

This challenge was further compounded by the nature of guidance they received. Assistance from companions typically emphasized movement, \Modified{offering direction but little situational context}. As P10 shared, \textit{``My mum just leads me—doesn’t tell me what’s around,''} highlighting how navigational support \Modified{rarely conveyed information about nearby features that could lead to exploratory decisions}. \Modified{Navigation apps reinforced this pattern. They focused on getting users to the destination efficiently, providing prescriptive turn-by-turn instructions without surfacing nearby points of interest or alternative paths, leaving little opportunity for unexpected or self-initiated discovery.}

\Modified{These limitations stood in stark contrast to participants’ unmet curiosity and desire for real-time environmental details that would help them make sense of, and actively explore, what was immediately around them, such as plant identity, surface characteristics, or the structure of nearby features.}
P4 emphasized, \textit{``Sighted people rely on the small signs in flowerbeds to know what the plants are. Since we cannot see them, we need such explanations even more.''} \Modified{However, such information was rarely available on-site, leaving participants with curiosity but without the informational affordances needed to turn that curiosity into exploration}.

Taken together, these experiences point to a broader gap between the desire for active, informed exploration and the lack of accessible structures to support it. \Modified{The barrier here lies in a disrupted exploratory mechanism: without \textbf{contextual cues and situational prompts}, exploration collapses into guided movement rather than active discovery.} 

\subsection{\Modified{Multisensory Strategies for Environmental Understanding}}
\label{Multisensory Approaches in Nature Engagement}

While barriers such as fragmented information and passive guidance constrained active exploration, participants described how they actively constructed their engagement with nature through multiple senses—hearing, smell, and touch. \Modified{These senses supported orientation, enriched environmental understanding, and enabled forms of exploration that compensated for the lack of visual access.}

\Modified{Hearing was the most frequently emphasized and was regarded as foundational for both mobility and situational awareness. Participants emphasized how they relied on continuous auditory cues to understand spatial layout, detect nearby activity, and identify potential hazards}. Quiet soundscapes helped them maintain direction, while noise or obstructed ears disrupted orientation. As P15 explained, \textit{``I prefer quiet places. Since we rely on hearing, when it gets noisy, you lose your sense of direction.''} \Modified{Beyond navigation, natural sounds also conveyed detailed environmental information such as animal presence and subtle natural cues like wind, flowing water, or falling leaves, that helped participants infer what was happening around them.}
P5 noted, \textit{``In autumn, when I go out early in the morning, I hear many different bird calls, and also insects like crickets in the green areas.''} 

\textbf{Smell} \Modified{served as another meaningful source of environmental information. It signaled seasonal transitions, vegetation changes, and local conditions. Participants described how olfactory cues made such shifts tangible}. P5 remarked, \textit{``When the grass begins to sprout, you can smell the freshness of plants and the earthy scent of soil.''} \Modified{Floral scents also allowed participants to infer characteristics such as plant density or bloom stage.} P6 shared, \textit{``When flowers are in full bloom, as soon as you smell the fragrance it almost feels as if you can see a sea of flowers.''}.

\textbf{Touch} enabled physical connection with natural elements, \Modified{providing concrete information that other senses could not. Participants described using touch to probe textures, temperatures, and physical structures}. P7 described, \textit{``I dip my hand into the water, feel its temperature, feel its touch...''} P11 recalled, \textit{``I like to touch flowers and feel their soft texture.''} 

Across these senses, participants expressed personal hierarchies. Hearing was most commonly prioritized and often described as essential for mobility. Others, however, emphasized the value of smell or touch. P15 said, \textit{``Hearing comes first, smell second... If I can’t hear, I can’t walk.''} P11 shared, \textit{``Smell is very helpful to me... I really depend on it.''} In contrast, P12 insisted, \textit{``I rely most on touch, because it tells you what something actually is.''} These differences reflected both personal preference and the demands of specific environments.

In sum, blind people engage with nature through a \textbf{multisensory system} in which hearing supports orientation and situational awareness, smell signals seasonal and spatial changes, and touch enables direct material contact. These senses are not used in isolation but are flexibly prioritized according to personal preference, context, and purpose.

\section{\Modified{Findings: Strong Emotional Connectedness to Nature with Constrained Ecological Awareness}}
\label{attitudes}
\Modified{This section shows how participants made sense of their relationship with nature across three dimensions. Emotionally, nature offered calm, vitality, and a sense of companionship. Attitudinally, those emotional attachments translated into sustained efforts to stay connected, which were expressed through personal routines and participation in organized outdoor activities. Ecologically, however, participants' ecological awareness was limited by their restricted access to environmental information, leaving them aware of ecological concerns yet uncertain about how to meaningfully engage with them. Together, these dimensions highlight how emotional grounding, sustained practices, and constrained ecological knowledge shaped participants’ lived relationship with nature.}

\subsection{\Modified{Nature as a Source of Emotional Grounding and Connectedness}}
\label{Emotional Value}

For blind people, nature is not merely a backdrop but a vital emotional anchor in navigating a world shaped by sensory loss and social barriers. Natural environments served not only as sources of sensory richness but also as emotionally significant spaces that foster calm, provide comfort, and nurture belonging rarely found in built environments.

A prominent \Modified{pattern} across interviews was the \textbf{restorative calm and sense of safety} that nature provided through its stability and predictability. In contrast to the uncertainty and overstimulation of urban life, natural settings offered a calming rhythm that helped participants feel emotionally grounded and secure. As P8 recalled, \textit{``Sitting on a rock and listening to water was the most soothing and relaxing experience.''} 
For P12, the emotional comfort of familiar natural elements extended into daily life: \textit{``After summer rains in childhood, the sound of frogs by the pond always helped me sleep peacefully, and even now I play frog sounds at night for comfort.''} Similarly, P5 described how seasonal change fostered emotional equilibrium: \textit{``In spring, the freshness of grasses and soil reminded me of plants beginning to grow... This natural atmosphere created a sense of peace, allowing me to feel fully immersed and at ease.''} These reflections underscore how natural settings created spaces where our participants could temporarily suspend their need for vigilance and regain a sense of control. In settings that often felt inaccessible or indifferent, nature stood out as an emotionally reliable presence.

Beyond its calming effects, participants also highlighted nature as a source of \textbf{emotional vitality and relational connection}. These experiences were especially valuable for those navigating social isolation.
As P4 remarked, \textit{``Visiting parks or hiking lifted my spirits and filled me with delight, especially when accompanied by birdsong.''} For P9, liveliness itself became emotionally affirming: \textit{``Being near the lake gave me a sense of energy, because the smell of fish in the air told me there were fish swimming in the water.''} These encounters were not just refreshing—they affirmed participants’ connection to a living, responsive world. This connection extended beyond internal mood shifts into subtle, affective relationships with the environment itself. As P11 reflected, \textit{``Some birds chirp so vividly—it feels like they’re talking to me.''} For her and others, such \Modified{moments were interpreted as gentle acknowledgements that offered a sense of recognition and companionship}. \Modified{In the context of limited social interaction and reduced access to the surrounding world, nature’s responsiveness provided an emotional partner that eased feelings of solitude}.

\Modified{Overall, these accounts show that blind people experience nature not simply as a relaxing escape, but as a multifaceted emotional resource. It serves to ground them in restorative calm and safety, simultaneously lifting their mood with vitality and offering moments of recognition and companionship. This rich emotional provisioning helps offset the vigilance, uncertainty, and isolation often present in participants’ everyday lives, particularly in built environments.}

\subsection{\Modified{Emotional Attachment Fosters Enduring Commitment to Nature Engagement}}
\label{Attitudes Toward Nature Engagement}

The emotional significance of nature for blind people was not limited to temporary comfort. It also shaped their enduring attitudes and consistent efforts to engage with the natural world. Participants did not perceive nature as something inaccessible; rather, they described it as a valued presence that they intentionally maintained in their lives.

Nature was frequently described as deeply important to their sense of wellbeing and everyday orientation toward the world. P6, who lost sight later in life, reflected, \textit{``I value nature even more now than when I could see...''} This sentiment illustrates how blindness often increased the perceived significance of natural environments. P11 similarly remarked, \textit{``Being in nature makes us feel good. I don’t think people can live without it.''} These reflections reveal a sustained and deliberate orientation toward nature, rooted in long-term appreciation rather than visual experience.

Such orientation was reflected in participants’ everyday behaviors. P5 mentioned, \textit{``I love walking, especially long walks.''} P9 explained her seasonal habit of going out early in the morning: \textit{``In summer, I usually go out around five in the morning to walk or jog in different parks.''} For P12, time outdoors was a consistent practice: \textit{``Whenever I have time, I go out for a walk. It’s part of my everyday routine.''} These examples illustrate a form of \textbf{embodied routine}, where interaction with nature was integrated into daily life rather than treated as an occasional event. In addition to personal routines, five participants engaged in structured group activities to maintain outdoor presence. P9 shared, \textit{``I joined a national blind running group. We go running with volunteers using guiding ropes... I gradually fell in love with the activity.''} Similarly, P11 described, \textit{``We have a running group and go out with volunteers every Saturday.''} These cases show how organized participation provided both access and motivation, reinforcing a sustained connection with nature through social means.

However, not all participants were able to sustained active engagement with nature.
P7 noted that external obligations sometimes interfered with their engagement. He explained, \textit{``Work is the most important thing. I have to stay in the shop waiting for customers every day, so I don't have time to go out into nature alone.''} While less common, such an example demonstrates how structural constraints could limit the ability to act on one’s environmental commitments.

These accounts suggest that blind people’s relationship with nature was not passive or incidental. It was intentionally cultivated through \textbf{enduring appreciation}, \textbf{repetitive practice}, and \textbf{social engagement}, reflecting a clear commitment to remaining connected to natural environments.

\subsection{\Modified{Ecological Understanding Remains Limited and Difficult to Form}}
\label{Environmental Attitudes}
Beyond their personal routines and emotional ties to nature, participants also reflected on how they understood larger \textbf{environmental issues}. \Modified{In these accounts, they described having limited ways to access information about topics such as conservation, pollution, or ecological change in a direct and independent manner.}
P3 explained, \textit{``I do not have many channels to learn about environmental issues. Mostly it is from what others tell me, and I listen to some news, but it is limited.''} 
\Modified{P7 added that this reliance on second-hand accounts shaped how he related to environmental concerns}: \textit{``When others tell me animals are suffering, I sympathize, but if it is only through sounds, it is hard for me to imagine.''} \Modified{These reflections suggest that mainstream presentations of environmental topics, often dependent on visual media, left participants with ecological understanding that was fragmented, indirect, and difficult to consolidate.}

Concerns about their ability to contribute to environmental action were also frequently voiced. Even when participants expressed attention to environmental change, they felt unable to influence it. P12 remarked, \textit{``I do pay attention to environmental changes in daily life, but I feel powerless, because I can’t do anything.''} P16 echoed this sense of limitation: \textit{``When it comes to environmental protection, maybe I can do less than sighted people.''} \Modified{Such perspectives reveal how limited access to participation channels thus translated into reduced confidence in contributing to ecological efforts.}

Taken together, \Modified{these accounts show that while participants cared about environmental issues, their ecological awareness remained constrained by difficulties accessing and interpreting information independently, as well as by a diminished sense of being able to participate in environmental action}.

\section{\Modified{Findings: Expectations for Technologies That Align with Blind People’s Ways of Nature Engagement}}
\label{expectation}
This chapter outlines key expectations for assistive technologies that support blind people in engaging with natural environments across four domains: information access, immersive simulation, nature recording, and device deployment.

\subsection{Diverse and Personalized Strategies for Information Presentation}
\label{Information Presentation}
Our participants expressed highly diverse expectations toward information presentation in natural environments, shaped in part by whether they were congenitally blind or had lost vision later in life. Those with prior visual experience, such as P5 and P10, often preferred concise, label-style descriptions. P10 remarked, \textit{``I don’t need everything explained. I just want to know what kind of tree it is. If I’m curious, I’ll ask for more.''} This reflects a demand for efficiency and control, allowing users to quickly identify elements while retaining the option for deeper exploration. In contrast, congenitally blind participants such as P9, P11, and P6, often needed rich, vivid descriptions to build foundational spatial or conceptual understanding. P9 explained, \textit{``Some people don’t even have the concept of `square'... You describe it, and they start to imagine.''} P11 echoed this with a more affective tone: \textit{``I think about the birds—how far apart they are, what color their feathers are, if their tails flick up when they call… I imagine it all.''} These responses highlight a form of imaginative engagement that depends on detailed verbal scaffolding to compensate for the lack of visual memory.
At the same time, two participants voiced a preference for emotionally expressive narration over synthetic text-to-speech. P5 stated simply, \textit{``If it’s just mechanical voice, I won’t listen. I prefer real human narration—it feels more alive.''}

However, current assistive technologies rarely account for these nuanced needs. Most systems are built for generic navigation and do not differentiate between user types, modes of interaction, or emotional tone. As P14 noted, \textit{``They either don’t read what they should or just read too much and confuse us.''} These accounts highlight that blind people’ expectations for information are not homogeneous, but shaped by blindness history and preferences for emotional resonance. A more nuanced understanding of these diverse strategies is essential for supporting rich and situated engagement with nature.

\subsection{Expectations for Nature Simulation and Immersive Experiences}
\label{Nature Simulation and Immersive Experiences}

Due to mobility challenges, safety concerns, and limited access to natural spaces, nine participants expressed strong interest in indoor simulations that evoke the emotional and atmospheric qualities of nature. As P7 noted, \textit{“Sometimes it's just too difficult to go far or find someone to guide me—I can't get to nature even if I want to.”} In these situations, simulations were not seen as mere substitutes, but as valuable avenues for reconnecting with natural elements. P15 shared, \textit{“Just sitting and listening to nature sounds can bring comfort... It helps me clear my mind and feel at ease.”} Rather than seeking exact replication, participants emphasized the importance of a \textbf{multisensory composition} that could foster both immersion and emotional resonance.

\textbf{Auditory layers} were viewed as the most crucial element. P7 envisioned a space filled with overlapping sounds of pine winds, ocean waves, and seagull calls, while P10 emphasized that \textit{``Music and sound affect people the most—especially when you don’t rely on sight.''} P9 described how subtle layers of animal sounds helped her imagine a dense forest: \textit{``You only feel like you’re in nature when you can hear all those small, staggered sounds.''} \textbf{Tactile elements} were also important. P12 emphasized that \textit{``Real tactile feedback makes immersion more believable.''} Similarly, P6 and P11 stressed that temperature and airflow enhanced their sense of being in a natural space. For instance, P11 imagined sitting on a grassy plain, \textit{``feeling the wind and hearing sheep nearby.''} \textbf{Smell} was mentioned by three participants (P10, P12, P15) as enriching the atmosphere, especially scents of flowers, wet soil, or rain-soaked forests. These layers helped trigger emotional responses, such as relaxation and joy.

When it came to \textbf{movement}, seven participants preferred real walking over virtual transitions. P6 noted, \textit{``Walking and physically feeling things is more real.''} P5 emphasized the value of bodily control: \textit{``When I walk, I move with my own rhythm—I can pause, slow down, or stop when I feel like it. Even virtual movement should let me set the pace, not follow a fixed playback.''} In contrast, P10 preferred virtual movement to avoid breaking immersion: \textit{``When I bump into something, I remember I’m still blind. It ruins the dream.''}

Participants envisioned nature simulation not as technical mimicry, but as a thoughtfully orchestrated immersive atmosphere built through sound, touch, smell, airflow, and movement, to evoke a sense of natural presence, support emotional regulation, and enable rich engagement when direct access to nature is constrained.

\subsection{Nature Photography for Recording and Memory}
\label{Recording and Social Sharing}

As P12 emphasized, \textit{``A photo is both a memory and a tool for expression. That’s important...''} For our participants, photography was not merely a technical act, but an important way to express identity, convey presence, and create emotional resonance.

However, this expressive motivation was often interwoven with \textbf{social expectations}. As P10 revealed, \textit{``I don't want people to know my eyes are bad... I don't want to explain...''} This concern about being identified as blind led participants to place greater emphasis on producing high-quality images. The desire to avoid unwanted attention or pity became a strong, if unspoken, driver of photographic effort. In this context, \textbf{aesthetic quality} was more than an artistic preference; it became a form of self-protection and social alignment. As P7 remarked, \textit{``I want my photos to look beautiful to others...''} Participants hoped their photos would not only be expressive, but also indistinguishable in quality from those taken by sighted individuals. As such, the act of taking a “good photo” often served to assert parity and agency in a visual culture.

Yet the process of capturing such photos posed major challenges. Without visual feedback, participants struggled with framing, composition, and physical orientation. As P5 asked, \textit{``How can I take a photo if I can't see...''}, pointing out the conceptual and practical gaps in conventional photographic tools. P6 added, \textit{``I need guidance on how to hold the phone at the best angle...''}, underscoring the need for \textbf{accessible, real-time feedback} during photo capture.

Even when a photo was successfully taken, its role as a \textbf{memory artifact} was often undermined by the difficulty of interpreting it afterwards. As P15 explained, \textit{``I don't know what's in the photo...''}, indicating that the photo’s emotional and narrative significance could be lost without contextual anchors. To address this, P5 proposed, \textit{``Automatic text descriptions and archiving would be great...''}.

Altogether, participants’ experiences reveal that nature photography is not simply about capturing an image. It is a layered, emotionally loaded act, serving as a means of self-expression, a response to societal perception, and a tool for emotional recall. Supporting blind people in documenting and remembering nature thus requires inclusive design that goes beyond access, enabling authorship, beauty, and lasting resonance.

\subsection{Form and Placement Preferences for Outdoor Assistive Devices}
\label{tech_preferences_outdoor}

Nine participants expressed clear needs for assistive technologies that support independent navigation and orientation in natural outdoor environments. Such systems were expected to be flexible, personalizable, and adaptable, functioning effectively across variable terrain, weather, and sound conditions without compromising mobility.

Seven participants favored \textbf{portable or wearable devices}, citing ease of use and integration with daily movement. P10 described a preferred setup as \textit{``a phone with Bluetooth earphones—not a heavy or weird-looking headset.''} P15 added that handheld devices were often impractical: \textit{``It’s better if it’s small, light, and worn on the body.''} Participants also emphasized what portable devices should avoid: visually stigmatizing appearances, complex controls, or interference with multitasking. P11 appreciated being able to use and share such systems flexibly across home and outdoor contexts.
Five participants preferred \textbf{site-specific installations} in public parks or trails, especially for users without access to personal devices. These setups could include stationary speakers, tactile markers, or descriptive sound stations. As P12 noted, \textit{``Many blind people can’t afford high-tech gear. Shared installations are better, or tools that run on phones.''}
Three participants supported hybrid approaches such as fixed audio posts that link to mobile devices or spatially distributed sound trails. These options offered a compromise between immersion and flexibility, especially along long paths or multi-user environments.

In addition, four participants raised concerns that assistive infrastructure might disrupt the \textbf{authenticity of culturally significant nature sites}. They feared that permanent, overly technical features would undermine the atmosphere of locations valued for their natural and historical qualities. As P16 warned, \textit{``If you add elevators to the Great Wall, is it still the Great Wall? Some places should stay natural.''}

These findings point to a need for assistive systems that are both inclusive and environmentally respectful, and should prioritize minimal, context-sensitive interventions that balance access, usability, and the integrity of natural experience.

\section{Discussion}  
\label{discussion}
In this section, we reflect on our findings to propose design implications for supporting blind people’s engagement with nature. The survey indicated lower levels of nature relatedness among blind participants. The interviews further highlighted that engagement was shaped by multisensory strategies within safe and familiar contexts, together with emotional connections and expectations for assistive technologies. Building on these insights, we outline design implications to inform technologies that foster safe, autonomous, and emotionally resonant forms of nature engagement for blind people.

\subsection{Balancing Safety and Autonomy in Nature Engagement Design}
\label{discussion_safety}
Our study reveals a central tension in blind people’s engagement with nature: safety is indispensable, yet excessive protective practices such as family restrictions or reliance on companions often undermined autonomy and curtailed opportunities for independent participation (see Section~\ref{safe}). \Modified{Autonomy is commonly defined as the capacity for self-governance or self-determination, meaning the ability to act according to a self-chosen plan or rule without controlling interference from others~\cite{kant2020groundwork}. Here, we use autonomy to refer specifically to blind people's ability to decide when, where, and how to move and explore in natural environments. It is important to note that the limitations on autonomy described by our participants did not arise from a lack of capability. Rather, these constraints were shaped by external and structural conditions such as family overprotection, inaccessible environments, and technologies that prioritize safety over user choice.} Outings were often limited to times when companions were available, activities were confined to familiar and nearby locations, and interactions with the environment were reduced to listening and smelling rather than tactile exploration or physical interaction. 

This tension has long been discussed in disability studies and care ethics~\cite{wolpert1980dignity, morris1991pride, kittay2001caring, mustaniemi2023vulnerability}. Wolpert’s notion of the “dignity of risk” warns that denying reasonable opportunities for risk-taking undermines personal development~\cite{wolpert1980dignity}, and Kittay argues that care relations, when overly protective, can suppress autonomy~\cite{kittay2001caring}. Such tensions also surface in assistive technologies like ultrasonic canes~\cite{panazan2024intelligent} and IoT smart sticks~\cite{farooq2022iot}, whose frequent alerts may discourage exploration. Mustaniemi-Laakso et al. argue that vulnerability does not preclude people’s ability to actively shape decisions and arrangements that affect their lives, provided that surrounding support structures enable such participation~\cite{mustaniemi2023vulnerability}. In the context of nature engagement, participants valued the assurance that safety provides, yet were frustrated by uniform protective approaches that restricted their ability to choose routes, explore unfamiliar areas, and engage independently with nature. This frustration reflected not a rejection of safety, but of diminished \Modified{personal autonomy}. Considering this, we believe there is a need to explore design approaches that balance safety with autonomy in the context of nature engagement. 

Effective design for nature engagement should build on safety as a baseline while creating opportunities to enhance user \Modified{autonomy}. Assistive systems should provide adjustable support based on context and user confidence, such as offering minimal cues on familiar paths and increasing guidance in unfamiliar terrain. Family anxiety is another key barrier to independence, yet often overlooked in design. Features like automatic check-ins at entry or return points can reassure caregivers without requiring constant supervision. This can help reduce overprotection and support user \Modified{autonomy}.

\subsection{Supporting Serendipitous Nature Discovery beyond Goal-Oriented Navigation}
\label{discussion_serendipity}
From our interviews, we learned that blind people’s engagement with nature is often shaped by goal-oriented navigation, which prioritizes reaching a destination but limits opportunities to notice and explore natural elements along the way (see Section~\ref{Active Exploration}). Guidance from companions or digital tools focused on efficiency, with little attention to features such as flowers, birds, or trees. However, participants emphasized that nature was not only about arriving somewhere, but also about the experience of being along the way.

Existing navigation systems typically offer limited support for exploration and contextual awareness~\cite{ahmetovic2016navcog,sato2017navcog3}. Recent work has begun to address this: systems like \textit{WanderGuide} and \textit{ObjectFinder} move beyond turn-by-turn instructions by enabling flexible, user-led access to environmental details~\cite{kuribayashi2025wanderguide, liu2024objectfinder, li2023understanding}. Our study extends this conversation to nature, where exploration is not just a means but a goal. Unlike urban or indoor spaces, nature requires support for unstructured, serendipitous discovery, where value lies in being in and with nature, not simply reaching a destination.

From this perspective, we identify two core needs based on our interview data (see Section~\ref{Active Exploration}): support for spontaneously noticing and engaging with nearby natural elements, and support for making sense of them by understanding what they are, how they behave, or how they fit within the environment. Designing for these needs means supporting spontaneous discovery alongside navigation. Future systems should help blind people detect and interpret nearby natural elements through spatialized audio or tactile cues, encouraging multisensory interaction such as touching textures or noticing scents. These features should be adaptable, allowing users to shift between direct travel and exploratory modes. Ultimately, assistive technologies for nature should act as companions that foster serendipitous, autonomous engagement.

\subsection{Empowering Environmental Agency Among Blind People}
\label{discussion_agency}
Our study found that blind people’s relationship with nature often remains one-directional, centered on emotional reception rather than mutual care and responsibility (see Section~\ref{Attitudes Toward Nature Engagement}; Section~\ref{Environmental Attitudes}). Our participants described two main barriers to environmental engagement: limited access to reliable information and a sense of helplessness about making a difference. This was also reflected in our survey, where blind people scored lower on environmental responsibility and concern for non-human life. \Modified{These barriers constrained what we describe in this paper as environmental agency. Here, we use environmental agency to refer to the capacity to understand environmental issues, develop ecological awareness, and take meaningful action in environmental care~\cite{rickinson2001learners}.}

One factor behind this gap is limited access to environmental education. In many contexts, especially in developing regions, blind people have fewer opportunities for formal learning, which constrains both ecological knowledge and confidence~\cite{world2019world, sen2014development, bani2024barriers}. Environmental discourse and citizen participation are also shaped by visual media, printed campaigns, and sighted-centered group activities, which often exclude them from public conversations and collective action. Prior studies have shown that education and participation play critical roles in fostering ecological responsibility~\cite{van2022does, newman2012future, silvertown2009new}, yet accessibility remains overlooked. Our findings highlight the need for inclusive pathways that extend emotional connection into awareness and action.

For design, this calls for participatory strategies that support blind people’s environmental agency. Eco-education can use auditory storytelling, tactile exploration, or group discussion to communicate ecological processes. Sustainability practices such as community gardening or sound-guided recycling can embed responsibility into shared routines. Citizen science projects may also involve blind people in recording bird calls, monitoring seasonal change, or keeping audio journals, positioning them as active contributors to environmental care.

\subsection{Designing Digital Nature as a Complementary Pathway to Nature Engagement}
\label{discussion_digitalnature}
Our findings show that although blind people value direct encounters with nature, social infrastructure, and mobility barriers often make spontaneous access difficult (see Section~\ref{experiences and challenges}). Therefore, our participants expressed interest in the digital nature that could supplement limited opportunities for physical contact (see Section~\ref{Nature Simulation and Immersive Experiences}).

In interviews, they described combining hearing, smell, and touch to create coherent and emotionally rich encounters (see Section~\ref{Nature Simulation and Immersive Experiences}). Hearing supported atmosphere, smell conveyed seasonal cues, and touch grounded experiences in material presence. These senses worked together to create affective qualities such as calm and companionship. Reflecting these practices, participants expected the digital nature not only to substitute for access but also to support emotional resonance. Prior HCI research has explored digital nature for relaxation and restoration~\cite{ma2025elements, spangenberger2024embodying}, and highlighted multisensory integration in shaping affective experience~\cite{obrist2025multisensory, rao2025designing}. However, most studies focus on general users, with little attention to the needs of blind people, which differ in sensory access and interaction priorities. Studies with blind people demonstrate the potential for multisensory interaction, including audio-tactile maps~\cite{brock2015interactive}, ultrasonic haptic feedback~\cite{tivadar2023learning}, and reviews stressing that design must balance user experience with sensory compensation~\cite{arora2024comprehensive}. Building on these threads, our study extends the discussion to nature engagement, emphasizing digital nature as a medium that fosters emotional depth while balancing user experience with sensory compensation.

For design, this implies creating interactive environments that invite engagement while avoiding sensory overload through adjustable auditory, tactile, and olfactory feedback. Soundscapes can layer birdsong, water, or wind to shape the atmosphere with controllable intensity. Tactile interfaces can complement other modalities through textures or symbolic forms. Olfactory cues can serve as subtle ecological markers, calibrated to harmonize with auditory and tactile input. Together, these strategies highlight opportunities for the digital nature to go beyond functional access and support participatory experiences that are emotionally resonant.

\subsection{\Modified{Personalizing Nature Descriptions for In-Situ Engagement}}
\label{discussion_information}
As shown in our findings, our participants have different expectations about \Modified{how nature-related information should be delivered for real-time interaction}, closely tied to their blindness history (see Section~\ref{Information Presentation}). Participants with prior visual experience preferred concise, label-like descriptions that activated existing visual images or spatial models. In contrast, congenitally blind participants sought more vivid narration to scaffold spatial understanding and imagination. \Modified{This highlights the need for flexible information presentation that allows users to control the method of description based on their individual blindness history, ensuring utility during current engagement}.

These insights also reveal limitations in current assistive systems, particularly in navigation and object recognition, which often adopt a “one-size-fits-all” approach. Prior work highlights ability-based design~\cite{wobbrock2011ability}, personalized mobility cues in navigation systems~\cite{ohnbar2018personalized}, and “softerware” as user-customizable interfaces that can be tuned to diverse needs and contexts~\cite{elavsky2025towards}. \Modified{Our study shows that the chosen method of description significantly affects users' ability to imagine scenes and feel emotional resonance, which are crucial for effective in-situ nature engagement.}

For design, this suggests that both the \Modified{description method used in real-time systems} should be personalized based on factors such as blindness history and individual preferences. For instance, \Modified{a label-based method, such as “a patch of daisies on your left,”} may suffice for users with prior visual experience, while congenitally blind users might benefit from \Modified{a narration method that scaffolds understanding}: “on your left, soft white petals circle a raised yellow center, about the size of a coin, growing close to the ground.” Initial settings can be guided by the blindness history, and adaptive models can refine the output over time. In sum, systems should shift from \Modified{a static, one-size-fits-all delivery to personalized description methods,} helping blind users access nature-related content in ways that are both efficient and experientially rich.

\subsection{\Modified{Documenting Nature to Preserve Experiences and Support Future Memory Recall}}
\label{discussion_memory}
\Modified{Participants in our study described documentation as essential for preserving their nature experiences and for supporting later recall (see Section~\ref{Recording and Social Sharing}). However, they also noted that existing tools provide limited support for creating records that effectively serve as memory cues.}

Although tools like audio prompts and framing guidance support image capture~\cite{jayant2011supporting, vazquez2014assisted}, they rarely address \Modified{how captured materials will function as long-term memory anchors}. Studies have explored metadata~\cite{adams2016blind}, sonic archives~\cite{dib2010sonic, duel2018supporting}, and tangible mementos~\cite{petrelli2017tangible} to aid memory, while Yoo et al.~\cite{yoo2024remembering} showed that sound can expressively anchor it. In nature contexts, \Modified{supporting the preservation and later recall of experiences} is especially important yet more difficult, as natural experiences are often fleeting and multisensory.

From a design perspective, this means that nature documentation should \Modified{be oriented first and foremost toward preserving experiences and supporting later recall. Effective documentation systems should therefore address both the capture and retrieval phases of the documentation process. To support effective preservation, systems could unobtrusively capture and select meaningful moments, rather than requiring users to actively frame every shot. For example, using AI-supported cameras that automatically detect and store salient natural scenes or soundscapes during a walk. After capture, models such as NIMA~\cite{talebi2018nima} could help identify high-quality photos or suggest adjustments to improve their composition, ensuring the preserved record is meaningful. To support memory-making and later recall, systems must enrich each entry with multimodal and semantic metadata. Models like CLIP~\cite{radford2021clip} could associate images with textual labels and themes, and support text–image search, so that users can later retrieve entries by asking for particular kinds of scenes or feelings (e.g., “photos taken near water” or “moments that felt peaceful”).} Furthermore, to support memory-making, systems might embed multimodal metadata, such as semantic labels, emotional tags, and ambient sounds, into each entry. For example, a nature diary might combine ambient sounds and spoken reflections to support multisensory recall. Together, these features can help blind users document nature in ways that feel both personal and complete.

\subsection{Designing Harmonious Assistive Technologies for Nature Engagement}
\label{discussion_authenticity}
Our participants preferred assistive technologies in nature to be seamlessly integrated into the environment, providing support without disrupting the integrity of the experience (see Section~\ref{tech_preferences_outdoor}). This concern echoes heritage research, which stresses that digital interventions must respect the authenticity of sites rather than overshadow them~\cite{hirsch2024human}. Extending this perspective to nature engagement, our study highlights that assistive technologies are most effective when conceived as part of nature, not as external intrusions.

From a design standpoint, assistive technologies that blend into nature tend to incorporate both material integration (e.g., form, texture, and placement) and experiential harmony (e.g., sensory tone and rhythm). Their forms are expected to remain subtle, ecologically considerate, and complementary to natural elements, avoiding visible alteration of the landscape. Based on our data, we identify four design directions. First, assistive cues can draw on environmental features across modalities. Natural sounds such as birdsong or flowing water may be amplified as auditory anchors, while tactile guidance may use wood or stone with distinct surface patterns, preserving continuity with the landscape. Second, interventions can adopt sustainable forms, such as biodegradable paths or recyclable markers that leave minimal traces and can be deployed seasonally or temporarily without altering the authenticity of the landscape. Third, interaction design can align assistive information with the temporal rhythms of nature. Because natural environments shift across seasons, times of day, and animal activity patterns, human experience of nature is similarly time-bound. Technologies that adapt to these dynamics, such as foregrounding birdsong in the morning or emphasizing wind patterns in the evening, can make interaction feel like part of the living rhythm of nature rather than a static overlay. \Modified{Finally, for audio-based or mobile systems, designs should ensure that assistive output does not mask the natural soundscape. Given that environmental sounds are central to immersion for blind people, using open-back or bone-conduction headphones can deliver navigational or narrative information while preserving access to ambient auditory cues.}

In sum, assistive technologies in nature can act as integrated mediators, enabling technological support and natural experience to coexist while preserving authenticity and expanding accessibility for blind people.

\section{Limitation and Future work}
This study has several limitations. First, all participants were blind (legally or totally), and the work did not capture the experiences of people with low vision. Individuals with residual sight may employ multisensory strategies—combining peripheral or degraded vision with auditory, tactile, or olfactory inputs—to interpret environments. Second, the study was conducted in mainland China, which may limit the cultural generalizability of the findings. Perceptions and practices of nature engagement could differ in other sociocultural and environmental contexts. These factors limit the generalizability of our findings. Third, the NRS survey sample (N=20) was relatively small, preventing systematic validation of the quantitative results. In addition, the NRS was not originally designed for blind populations and may not fully capture non-visual pathways of nature connectedness.

Future research should address these limitations while extending our findings in new directions. Methodologically, more inclusive instruments of nature connectedness are needed that capture multisensory and social pathways rather than relying on vision-oriented constructs. Empirically, comparative studies could examine how people with low vision engage with nature, as well as how different cultural contexts and occupational groups shape ecological meaning-making. 
Larger-scale mixed-methods studies could further validate the relationships between questionnaire scores and lived experiences. \Modified{In addition, our analysis did not include blind researchers; future work could include blind researchers in coding and theme development, bringing additional experiential perspectives.} Beyond methodological extensions, future work could also explore how emerging technologies might support not only mobility but also the emotional and social dimensions of engagement, and assess their long-term impact on well-being, participation, and environmental attitudes in real-world natural settings. \Modified{We also call for assistive technologies to be developed through co-design with blind people, involving them throughout the design process to accurately meet their specific needs for nature engagement.}

\section{Conclusion}
In this paper, we conducted a comparative survey with 20 blind and 20 sighted participants and in-depth interviews with 16 blind participants to understand blind people’s ways of engaging with nature and the barriers and supports that influence them. We show that blind people engage with nature through familiar and socially supported routines, multisensory strategies for orientation and experience (e.g., birdsong, ground textures, seasonal scents), and perceptions of nature as a source of calm, meaning, and environmental connection. We also explore blind participants’ expectations for assistive technologies that go beyond mobility to support autonomy, multisensory inclusion, and social connection. Our work extends research on accessibility and human–nature interaction by contributing design insights specifically related to nature engagement. In turn, we propose future directions for inclusive, context-sensitive technologies that create richer and more equitable opportunities for blind people to connect with the natural world.

\begin{acks}
This work is supported by the National Natural Science Foundation of China (Grant No. 62302094). Juxiao Zhang acknowledges the support from the Research Projects of National Language Committee (No.YB145-120 Research on New Requirements for the Construction of Language and Script Standardization in Key Fields).
\end{acks}
\bibliographystyle{ACM-Reference-Format}
\bibliography{main}

\clearpage        
\onecolumn        
\appendix

\setcounter{figure}{0}   
\setcounter{table}{0}    

\section{Item-Level Results of the NRS-21}

\begin{figure}[H]
    \centering
    \includegraphics[width=\linewidth, trim=60 260 50 80, clip]{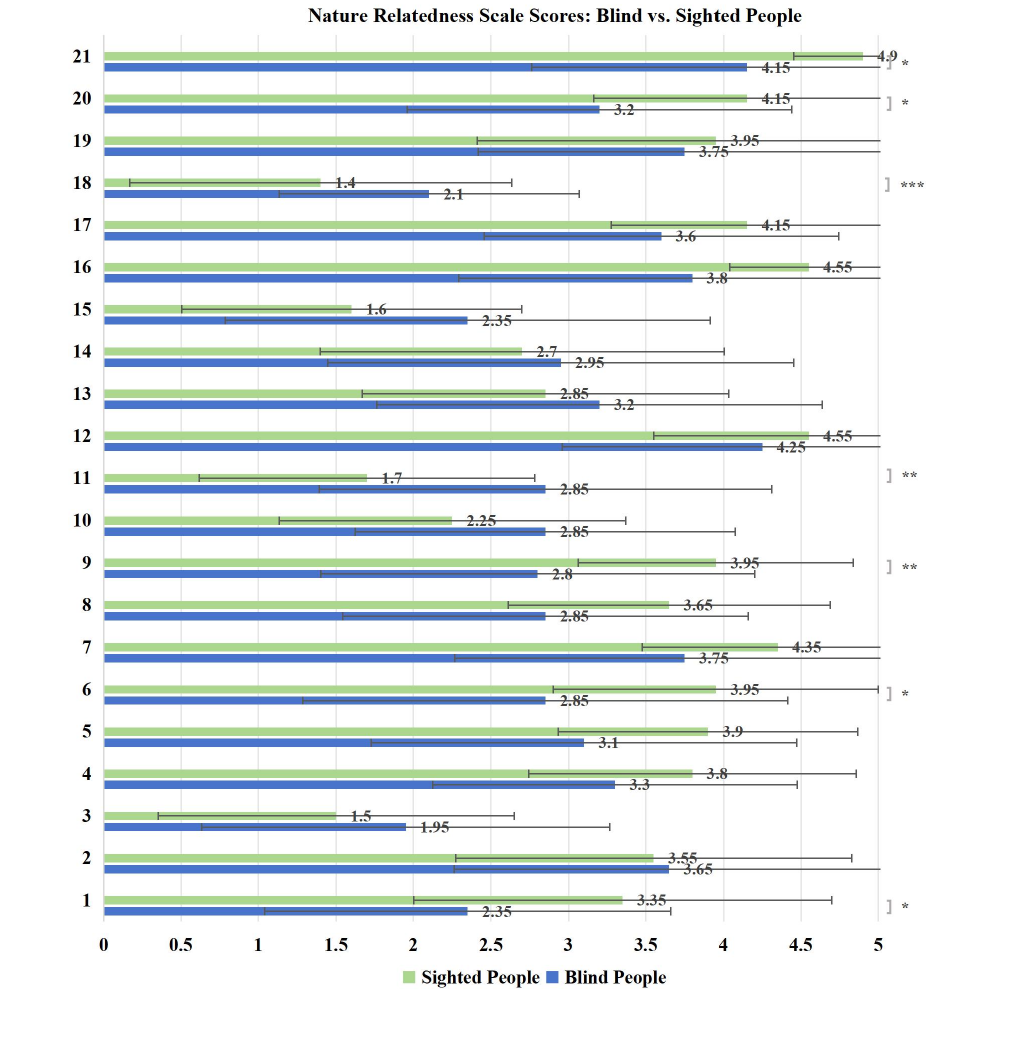}
    \caption{Item-level mean scores for the 21 Nature Relatedness Scale (NRS) items, comparing blind and sighted participants. Error bars show standard deviations. Asterisks indicate significant group differences (Mann–Whitney U tests; * $p < .05$, ** $p < .01$, *** $p < .001$). Item numbers correspond to Appendix A, Table~\ref{tab:nrs_items}.}
    \label{fig:nrs_appendix}
    \Description{Bar chart showing mean scores of blind and sighted participants across all 21 NRS items. Sighted participants consistently scored higher, though the size of differences varied by item. Error bars indicate variability (mean $\pm$ SD). Statistically significant group differences are marked by asterisks: one asterisk for $p < .05$, two for $p < .01$, and three for $p < .001$.}
\end{figure}

\newpage

\begin{table}[h]
\caption{Item-level statistical comparison of blind and sighted participants on the 21 NRS items. Mann--Whitney $U$ tests were conducted for each item, with $p$-values and effect sizes ($r$) reported.}
\Description{This table reports item-level comparisons between blind and sighted participants on the 21 items of the Nature Relatedness Scale, using Mann--Whitney U tests. Several items showed significant group differences, especially in NR-Experience and NR-Self dimensions. Blind participants reported lower outdoor enjoyment, reduced tactile engagement, and fewer notices of wildlife. They also showed lower environmental agency and empathy, for example agreeing more that conservation is unnecessary and reporting less concern for animals. Overall, the item-level results confirm that blind participants consistently scored lower in experiential, self-identification, and perspective-related aspects of nature connectedness.}
\label{tab:nrs_items}
\centering
\renewcommand{\arraystretch}{1.3}
\setlength{\tabcolsep}{3pt}
\begin{tabular}{|c|p{6.8cm}|c|c|c|c|}
\hline
\textbf{Item} & \textbf{Content} & \textbf{Dimension} & \textbf{$U$} & \textbf{$p$-value} & \textbf{$r$ (effect size)} \\
\hline
1  & I enjoy being outdoors, even in unpleasant weather. & NR-Experience   & 118 & 0.025  & 0.35 \\
\hline
2  & Some species are just meant to die out or become extinct. & NR-Perspective  & 187 & 0.724  & 0.06 \\
\hline
3  & Humans have the right to use natural resources any way we want. & NR-Perspective  & 154 & 0.134  & 0.20 \\
\hline
4  & My ideal vacation spot would be a remote, wilderness area. & NR-Experience   & 151 & 0.173  & 0.21 \\
\hline
5  & I always think about how my actions affect the environment. & NR-Self         & 133 & 0.061  & 0.29 \\
\hline
6  & I enjoy digging in the earth and getting dirt on my hands. & NR-Experience   & 115 & 0.018  & 0.36 \\
\hline
7  & My connection to nature and the environment is a part of my spirituality. & NR-Self         & 165 & 0.303  & 0.15 \\
\hline
8  & I am very aware of environmental issues. & NR-Self         & 131 & 0.057  & 0.29 \\
\hline
9  & I take notice of wildlife wherever I am. & NR-Experience   & 105 & 0.008  & 0.41 \\
\hline
10 & I don’t often go out in nature. & NR-Experience   & 141 & 0.102  & 0.25 \\
\hline
11 & Nothing I do will change problems in other places on the planet. & NR-Perspective  & 107 & 0.009  & 0.40 \\
\hline
12 & I am not separate from nature, but a part of nature. & NR-Self         & 177 & 0.451  & 0.10 \\
\hline
13 & The thought of being deep in the woods, away from civilization, is frightening. & NR-Experience   & 176 & 0.515  & 0.10 \\
\hline
14 & My feelings about nature do not affect how I live my life. & NR-Self         & 184 & 0.657  & 0.07 \\
\hline
15 & Animals, birds and plants have fewer rights than humans. & NR-Perspective  & 140 & 0.074  & 0.26 \\
\hline
16 & Even in the middle of the city, I notice nature around me. & NR-Self         & 163 & 0.276  & 0.16 \\
\hline
17 & My relationship to nature is an important part of who I am. & NR-Self         & 146 & 0.121  & 0.23 \\
\hline
18 & Conservation is unnecessary because nature is strong enough to recover from any human impact. & NR-Perspective  &  75 & $<.001$ & 0.54 \\
\hline
19 & The state of nonhuman species is an indicator of the future for humans. & NR-Perspective  & 172 & 0.415  & 0.12 \\
\hline
20 & I think a lot about the suffering of animals. & NR-Self  & 114 & 0.015  & 0.37 \\
\hline
21 & I feel very connected to all living things and the earth. & NR-Self         & 140 & 0.020  & 0.26 \\
\hline
\end{tabular}
\end{table}

\newpage
\section{\Modified{Interview Protocol}}

\Modified{All questions below were used flexibly depending on participants’ responses. The interview was organized into four sections, each focusing on a different aspect of participants’ experiences: (1) warm-up and demographic background, (2) everyday nature engagement and related challenges, (3) preferences and expectations regarding assistive technologies, and (4) broader attitudes toward nature and the human–nature relationship. This structure allowed us to move from concrete daily practices toward more reflective and value-oriented perspectives.}

\Modified{Except for the initial warm-up questions, which were asked first to establish rapport, the remaining sections followed a semi-structured format. The order of questions within each section was adapted dynamically based on participants’ narratives, allowing the interviewer to probe relevant themes while maintaining a natural conversational flow.}

\subsection*{\Modified{Warm-up and Demographics}}

\Modified{This section was designed to ease participants into the interview and establish a comfortable conversational atmosphere before moving into more experience-focused questions. The warm-up phase focused on non-sensitive, open-ended prompts about everyday life and surroundings, helping participants become familiar with the interview flow. Demographic questions were then introduced gradually, moving from general lifestyle and background to more personal topics such as visual condition or other health-related information. This sequencing aimed to minimize discomfort, build rapport, and support participants in sharing their experiences at their own pace.}

\Modified{We began with a brief introduction to the study, clarified the meaning of “nature” and “nature engagement,” and reminded participants that they could skip any question or pause the interview at any time.}

\begin{itemize}
    \item[1.] \Modified{Could you describe the place where you live and how you feel about the natural environment around it?}
    \item[2.] \Modified{What kinds of activities do you usually enjoy for relaxation or leisure?}
    \item[3.] \Modified{Could you tell me a bit about your current study or work situation?}
    \item[4.] \Modified{How would you describe your gender identity in the way you feel most comfortable?}
    \item[5.] \Modified{How would you describe your age or age group?}
    \item[6.] \Modified{In your own words, how would you describe your visual condition?}
    \item[7.] \Modified{Could you share how and when your visual impairment developed?}
    \item[8.] \Modified{Is there anything else about your physical or sensory health that you feel is relevant to your experiences?}
\end{itemize}

\subsection*{\Modified{Nature Engagement Experiences and Challenges}}

\Modified{This section focuses on participants’ everyday interactions with natural environments, including their routines, sensory strategies, and the challenges they encounter in different settings. The questions move from general engagement patterns toward more reflective descriptions of expectations, experiences, and barriers.}

\begin{itemize}

    \item[9.]  \Modified{Could you describe how often you usually engage with nature and what those moments typically look like?}
    \item[10.] \Modified{In your daily life, what kinds of activities or situations bring you into contact with nature?}

    \item[11.] \Modified{How do different conditions—such as weather or season—affect your willingness or comfort in engaging with nature?}
    \item[12.] \Modified{When you go into natural environments, how do you usually prefer to go—alone or with others—and what kinds of support feel helpful for you?}

    \item[13.] \Modified{In what ways do you share your nature experiences with others, if at all?}
    \item[14.] \Modified{Could you describe any tools or technologies you use when engaging with nature, and how they influence your experience?}

    \item[15.] \Modified{Could you share a particularly memorable or meaningful nature experience you’ve had?}
    \item[16.] \Modified{In urban settings, which natural elements—such as sounds, textures, or smells—help you relax or feel connected?}
    \item[17.] \Modified{How do your senses (e.g., sound, smell, touch) shape the way you understand or interact with natural environments?}

    \item[18.] \Modified{What challenges or difficulties have you encountered when trying to engage with nature in urban areas?}
    \item[19.] \Modified{How well do urban natural spaces match your personal expectations of “nature,” and in what ways do they differ?}

    \item[20.] \Modified{From your perspective, how does city nature compare with rural or more remote natural environments?}
    \item[21.] \Modified{Could you describe what motivates you—or holds you back—from seeking opportunities to engage with nature?}

\end{itemize}

\subsection*{\Modified{Preferences and Expectations for Assistive Technologies in Nature Engagement}}

\Modified{This section examines participants’ experiences with assistive technologies in natural settings and their expectations for future tools that might support nature engagement. The questions progress from past use toward preferred sensory feedback, device characteristics, and their impressions of virtual or simulated nature.}

\begin{itemize}

    \item[22.] \Modified{Could you describe any assistive technologies you have used while engaging with nature?}
    \item[23.] \Modified{How have these tools influenced your experiences, either positively or negatively?}

    \item[24.] \Modified{If future technologies were designed to support your connection with nature, what aspects would you like them to focus on?}
    \item[25.] \Modified{How do you feel about different forms of feedback—such as voice, vibration, or tactile cues—and in what situations do they feel most useful?}
    \item[26.] \Modified{What types of devices (e.g., portable, wearable, home-based) feel most comfortable or practical for your needs?}

    \item[27.] \Modified{Could you describe any experiences you’ve had with virtual or simulated nature, such as soundscapes or tactile effects?}
    \item[28.] \Modified{In what ways did those experiences feel realistic or different from being in real natural environments?}
    \item[29.] \Modified{How do you think virtual nature fits into your overall relationship with nature—what roles could it play for you?}
    \item[30.] \Modified{In what ways, if any, might virtual nature influence your motivation to experience real outdoor environments?}

    \item[31.] \Modified{If you could create a nature-like experience at home, what kinds of environments or elements would you choose?}
    \item[32.] \Modified{When imagining immersive systems like VR, how would you prefer to interact—through movement or through a seated, stationary experience?}
    \item[33.] \Modified{What personal outcomes would you hope to gain from such systems (e.g., relaxation, focus, emotional regulation, better sleep)?}

\end{itemize}

\subsection*{\Modified{Attitudes Toward Nature Engagement}}

\Modified{This section explores participants’ broader attitudes toward nature, including emotional connections, environmental awareness, and perceptions of the human–nature relationship. The questions gradually shift from personal feelings and meanings toward reflections on ecological values and the impact of vision loss on their engagement.}

\begin{itemize}

    \item[34.] \Modified{When you think about “nature,” how would you describe your personal sense of connection or distance to it?}
    \item[35.] \Modified{In what ways does nature contribute to your emotional or spiritual life?}

    \item[36.] \Modified{How do you notice or respond to changes in the natural environment in your daily routines?}
    \item[37.] \Modified{Are there moments in urban nature that have strengthened your sense of connection or belonging?}

    \item[38.] \Modified{How would you describe the relationship between humans and nature from your perspective?}
    \item[39.] \Modified{What roles do you think parks and other urban green spaces play in city life?}
    \item[40.] \Modified{How important do you feel nature protection is, especially in fast-developing cities?}

    \item[41.] \Modified{In what ways do you think your own actions or habits may influence the natural environment?}
    \item[42.] \Modified{How would your daily life be affected if plants or animals were significantly reduced or absent?}

    \item[43.] \Modified{Looking back, how has your engagement with nature changed over time, particularly before and after experiencing vision loss?}
    \item[44.] \Modified{How has vision loss shaped your motivation, comfort, or interest in engaging with natural environments?}

\end{itemize}

\end{document}